\documentclass[useAMS,usenatbib]{mn2e}
\usepackage{natbib}
\usepackage{epsf,psfig}
\usepackage{graphicx}
\usepackage{textcomp}

\def\simlt{\mathrel{\rlap{\lower 3pt\hbox{$\sim$}}
        \raise 2.0pt\hbox{$<$}}}
\def\simgt{\mathrel{\rlap{\lower 3pt\hbox{$\sim$}}
        \raise 2.0pt\hbox{$>$}}}

\title[On the powering mechanism of transient jets]{Observational constraints on the powering mechanism of transient relativistic jets}
\author[D. M. Russell et al.]{D. M. Russell$^{1,2}$\thanks{E-mail: russell@iac.es}, E. Gallo$^{3}$ \& R. P. Fender$^{4}$
\\
$^{1}$Instituto de Astrof\'isica de Canarias (IAC), E-38200 La Laguna, Tenerife, Spain\\
$^{2}$Departamento de Astrof\'isica, Universidad de La Laguna (ULL), E-38206 La Laguna, Tenerife, Spain\\
$^{3}$Department of Astronomy, University of Michigan, 500 Church Street, Ann Arbor, MI 48109, USA\\
$^{4}$School of Physics and Astronomy, University of Southampton, Highfield SO17 1BJ, UK\\
}
\begin{document}


\pagerange{\pageref{firstpage}--\pageref{lastpage}} \pubyear{2012}

\maketitle

\label{firstpage}

\begin{abstract}
We revisit the paradigm of the dependency of jet power on black hole spin in accreting black hole systems. In a previous paper we showed that the luminosity of compact jets continuously launched due to accretion onto black holes in X-ray binaries (analogous to those that dominate the kinetic feedback from AGN) do not appear to correlate with reported black hole spin measurements. It is therefore unclear whether extraction of the black hole spin energy is the main driver powering compact jets from accreting black holes. Occasionally, black hole X-ray binaries produce discrete, transient (ballistic) jets for a brief time over accretion state changes. Here, we quantify the dependence of the power of these transient jets (adopting two methods to infer the jet power) on black hole spin, making use of all the available data in the current literature, which includes 12 BHs with both measured spin parameters and radio flares over the state transition. In several sources, regular, well-sampled radio monitoring has shown that the peak radio flux differs dramatically depending on the outburst (up to a factor of 1000) whereas the total power required to energise the flare may only differ by a factor $\simlt 4$ between outbursts. The peak flux is determined by the total energy in the flare and the time over which it is radiated (which can vary considerably between outbursts). Using a Bayesian fitting routine we rule out a statistically significant positive correlation between transient jet power measured using these methods, and current estimates of black hole spin. Even when selecting subsamples of the data that disregard some methods of black hole spin measurement or jet power measurement, no correlation is found in all cases. \end{abstract}


\section{Introduction}

Accretion onto black holes commonly results in relativistic, collimated outflows, or jets. In black hole X-ray binaries (BHXBs), these jets can brighten, fade or even switch off all together on timescales of days--months \citep[see][for a review]{fend06}. When jets are observed in BHXBs they usually manifest themselves as powerful, compact outflows commonly seen as a flat or slightly inverted radio spectrum ($\alpha \sim 0$ -- 0.5, where $F_{\nu} \propto \nu^{\alpha}$) and are spatially resolved in some cases \citep*[e.g.][Miller-Jones et al. in preparation]{dhawet00,stiret01}. Both properties are in line with what is expected for partially self-absorbed overlapping synchrotron spectra \citep{blanko79}. These compact jets of BHXBs are typically launched continuously for several months at least during a single outburst, when the X-ray spectrum is a hard power law \citep[the `hard state'; e.g.][]{mcclre06,bell10} and are considered to be analogous to persistent compact radio sources associated with active galactic nuclei (AGN). In BHXBs, most of the energy radiated by these jets is emitted at higher energies, for example in the infrared/optical regime \citep*[e.g.][]{corbfe02,chatet11} or even higher -- X-ray or possibly $\gamma$-ray energies in some cases \citep*[e.g.][]{market01,russet10,lauret11}. The direct detection of a break in the jet spectrum in one BHXB, GX 339--4 at mid-infrared wavelengths has allowed a measurement of the total compact jet radiative power to be made, $> 6 \times 10^{35}$ erg s$^{-1}$ (assuming a distance of 8 kpc) which is $> 0.6$ per cent of the total bolometric X-ray luminosity \citep{gandet11}, \citep[and breaks have recently been constrained in several more BHXB jet spectra;][]{rahoet12,russet13}. These jets are radiatively inefficient, and the total (kinetic plus radiative) energy channelled into the jets can be approximately half of the available accretion energy or perhaps more \citep*{gallet05,russet07a,pakuet10}.

During accretion state transitions at high X-ray luminosities ($> 1$ per cent of the Eddington luminosity; $L_{\rm Edd}$), a different type of jet is commonly seen in BHXBs. The compact jet fades and discrete, ballistic ejections appear as bright, optically thin radio flares, sometimes later resolved into two expanding structures either side of the core BHXB \citep[e.g.][]{miraro94,tinget95}. These discrete ejections are typically seen once, or at most a few times during a BHXB outburst cycle. Each ejection has a total radiative luminosity of $\sim 10^{-2}$ -- 1 $L_{\rm Edd}$ \citep*[see e.g. table 1 in][]{fendet04}, and may be associated with a fast, relativistic ejection \citep*[see also][]{millet06}. For comparison, a compact jet such as that of GX 339--4 mentioned above produces this much radiative energy in $< 30$ minutes during a bright hard state \citep{gandet11}. Since hard states normally last several months, the transient jets probably provide a negligible fraction of the total radiative power of jets from BHXBs. However, for both types of jet, the total kinetic power they contain is largely unknown, so comparisons can only be made between their radiative luminosities. It is worth assessing whether the same physical process is responsible for launching both types of jet.

The physical mechanism responsible for the launching, acceleration and collimation of jets has been debated for decades, and remains one of the major paradigms in the field of accretion in strong gravitational fields. It is generally accepted that magnetohydrodynamic (MHD) processes are at work \citep*[e.g.][]{meieet01}, but two groups of models are usually considered that can both produce the poloidal magnetic fields and differential rotation necessary to launch powerful jets \citep{blanzn77,blanpa82}. The Blandford--Znajek mechanism describes the extraction of the rotational energy of spinning BHs, in a region close to the BH where frame-dragging occurs \citep*[$\simlt 100$ gravitational radii; $R_{\rm g} = G M/c^2$; e.g.][]{meieet01,mckiet12}. The alternative to spin-powered jets is purely accretion-powered jets; the Blandford--Payne mechanism, where all the energy and rotation originates in the disc and its magnetic fields. The Blandford--Znajek model predicts that the total jet power scales with the second power of the dimensionless spin parameter, $a_{\ast}=cJ/GM^2$, where $J$ and $M$ are the black hole angular momentum and mass \citep*[see][for a similar scaling based on general-relativistic magnetohydrodynamics simulations]{tcheet10}.

The two leading methods currently adopted to infer BH spin in BHXBs are X-ray thermal continuum fitting, and X-ray reflection/Fe K$_{\alpha}$ line fitting. The former applies to spectra of BHXBs in the thermal dominant (or high/soft) state \citep{mcclre06,bell10}, while the latter fits (low/)hard state spectra at Eddington rates higher than about $10^{-3}$. In essence, the thermal continuum fitting method provides a measure of the area subtended by the Innermost Stable Circular Orbit (ISCO), while the reflection method returns a measure of the gravitational redshift that shapes the red wing of the broadened Fe K$_{\alpha}$ line.  Both methods rely on the assumption that the accretion disc extends in to the ISCO, which is in turn a function of BH spin.  
The reader is referred to e.g. \cite{mill07} and \cite{mcclet11} for further details on the two methodologies. BHXBs, as opposed to AGN, are particularly suited for this kind of study. In outburst, Galactic BHXBs tend to be brighter than all AGN in the sky, so with more photons, spectral fitting can be achieved with a higher level of precision. In addition, the thermal continuum fitting method cannot be applied to AGN, since the peak of the disc emission falls in the UV band.

Measures of BH jet powers also rely on different methods, arguably all more indirect than those adopted for estimating the spin parameter, such as the relative normalization of the radio/infrared (IR) versus X-ray luminosity in the hard state for compact jets, and the total energy in the bright radio flares that are thought to be at the origin of the large scale, transient jets.

In a previous work \citep*[][hereafter FGR10]{fendet10} we performed the first study to observationally test for empirical relations between BH spin ($a_{\ast}$) and jet power. Radio--X-ray and IR--X-ray correlations exist in the hard state but some BHXBs are more radio-bright (or infrared-bright) than others, at a given X-ray luminosity. The luminosity of the jet at radio and IR frequencies relative to the X-ray luminosity (the normalization of the correlation) was used to rank the BHXBs in order of jet power. No single relation was found between current reported BH spin measurements and relative luminosity of compact jets. This suggested that if the relative jet power estimates and BH spin estimates are accurate, compact jets may not be spin-powered. 

Since the acceleration and collimation regions of compact jets are usually considered to be at least tens to hundreds of $R_{\rm g}$ \citep*[e.g.][]{junoet99,market05,gandet11,polket10,peerma12}, these results may not be surprising because this is essentially outside the region where BH spin energy can be tapped. In addition, the jet is usually required to be anchored to a thick disc, which is thought to be present in the hard X-ray state (e.g. \citealt{meie99,market05}; see also \citealt{polket10}; however a thin disc that extends close to the ISCO may exist above $10^{-3} L_{\rm Edd}$ in the hard state; \citealt*{reiset10}). Very recently, the actual base of the jet of the AGN in M87 has been spatially resolved using radio interferometry, and its size was measured to be $5.5 \pm 0.4 R_{\rm g}$ \citep{doelet12}. However even here, the wide opening angle of the sub-relativistic jet and the existence of a counter jet indicates that this jet is anchored within (and powered by) the accretion disk and not launched due to magnetic field lines crossing the event horizon \citep{doelet12}.

The radio/IR normalization for a single source can change in time \citep*{russet07b,coriet11,gallet12} which, if X-ray luminosity is not the quantity that is changing, invalidates a single relationship between radio/IR luminosity normalization and jet power (or jet radiative efficiency). Some MHD simulations of disc-emitting jets have also successfully described properties of the outflows commonly seen in BHXBs \citep[e.g.][]{ferret06,polket10}. In addition, \cite*{miglet11} found no significant correlation between the spin of neutron stars in X-ray binaries (which can be measured accurately for some but not all sources considered) and their relative jet powers inferred from radio or infrared jet luminosities (although less data were available for neutron star jets, and those data span a smaller range in luminosity than the sample of BHXBs). Finally, in a related study, \cite{bamb12} considers the Johannsen-Psaltis metric around a BH, a speculative, alternative space-time metric to the Kerr metric. In this scenario, current data of BHXBs suggest there may be a correlation between BH spin as defined by this metric, and compact jet power.

The only tentative relation found by FGR10 was between \emph{transient} jet ejection energy and BH spin measured via reflection fits (but not continuum fits), from four data points (fig. 6 in FGR10). More recently, \citeauthor{naramc12} (2012; hereafter NM12) claim a relation between transient jet peak radio luminosity and BH spin measured from thermal continuum fitting, using a sub-sample of the data in the literature. NM12 take the peak luminosity of the radio flare associated with the hard-to-soft state transition as their estimate of the jet power (as opposed to an estimate of the total energy of the ejection).

Here, we test for a relation between transient jet power and reported BH spin using all the available data (described in Section 2), adopting two differing methods to infer the jet power and two methods to estimate BH spin. We make use of a Bayesian fitting routine in order to quantify the significance of any relation (the methodology is explained in Section 3). In all cases, no significant correlation is found. If the methods used to infer jet power and BH spin are correct, we can rule out a single positive correlation between transient jet power and BH spin with high confidence. A discussion and brief summary is provided in Section 4. We also comment on a very recent paper, \cite*{steiet13} in Section 4. This new work follows on from that of NM12, adding one more source to their sample. This new source was already included in our samples here, and we demonstrate that this paper does not affect our results or conclusions.

\begin{table*}
\begin{center}
\caption{All BHXBs with both published radio jet flare data and BH spin estimates.}
\vspace{-0.5cm} 
\begin{tabular}{llllllllllll}
\hline
Source&$d$&$M_{\rm BH}$&\multicolumn{2}{c}{--------- BH spin $a_*$ ---------}&$S_{\rm 5GHz}$&$\Delta$t&Year&\multicolumn{4}{c}{-------------- References --------------}\\
       & (kpc) & ($M_\odot$) &disc                    &reflection        &(mJy) & (h)         &      & $d$ & $M_{\rm BH}$ & $a_*$ & $S_{\rm 5GHz}$ \\
\hline
A0620--00      & $1.06 \pm 0.12$ & $6.61 \pm 0.25$ & $0.12 \pm 0.19$        &                  &203   &             & 1975 & 1 & 1   & 2 & 3,4 \\
GRS 1124--68   & 5.1             & $6.0 \pm 1.5$   & -0.04$^a$              &                  &171   &             & 1991 & 5 & 6   & 7 & 8 \\
4U 1543--47    & $7.5 \pm 1.0$   & $9.4 \pm 1.0$   & $0.8 \pm 0.1$          & $0.3 \pm 0.1$    &17.5  & 8           & 2002 & 9 & 9   & 10,11 & 12 \\
XTE J1550--564 & $4.38 \pm 0.50$ & $9.10 \pm 0.61$ & $0.34 \pm 0.24$ &  $0.55^{+0.15}_{-0.22}$ &265   & 12          & 1998 & 13 & 13 & 14    & 15,4 \\
XTE J1652--453 & 8$^b$           & 10$^b$          &                        & $\sim 0.5$       &$>0.95$ &           & 2009 &--$^b$&--$^b$& 16 & 17 \\
GRO J1655--40  & $3.2 \pm 0.5$   & $6.30 \pm 0.27$ & $0.7 \pm 0.1$          & $0.98 \pm 0.01$  &2 420 & 12          & 1994 & 18 & 19 & 10,11 & 18,4 \\
               &                 &                 &                        &                  &3.6   & 71          & 2005 & & & & 20 \\
GX 339--4      & 8$^b$           & $12.3 \pm 1.4$  & $< 0.9$                & $0.94 \pm 0.02$  &55    & 5.5         & 2002 & 21$^b$ & 22 & 23,11 & 24 \\
H1743--322     & $8.5 \pm 0.8$   & $13.3 \pm 3.2$  & $0.2 \pm 0.3$          &                  &96.1  & $<$48       & 2003 & 25 & 22 & 25 & 26 \\
               &                 &                 &                        &                  &24.4  & $<$24       & 2009 & & & & 27 \\
XTE J1752--223 & $3.5 \pm 0.4$   & $9.5 \pm 0.9$   &                  & $0.52^{+0.16}_{-0.13}$ &20    &             & 2010 & 28 & 28 & 29 & 30 \\
GRS 1915+105   & $11.0 \pm 1.0$  & $14.0 \pm 4.4$  & $0.975 \pm 0.025$      & $0.98 \pm 0.01$  &912   & $<$119      & 1994 & 31 & 32 & 33,34 & 35,4 \\
               &                 &                 &                        &                  &356   & 12          & 1997 & & & & 31,4 \\
               &                 &                 &                        &                  &150   &             & 2006 & & & & 36 \\
               &                 &                 &                        &                  &356   & $<$24       & 2010 & & & & 37 \\
Cyg X--1       &$1.86^{+0.12}_{-0.11}$& $14.8 \pm 1.0$&$0.96 \pm 0.04$      & $0.97^{+0.014}_{-0.02}$ &186  & $>$0.3& 2005 & 38 & 39 & 40,41 & 42 \\
GS 2000+25     & $2.7 \pm 0.7$   & $7.2 \pm 1.7$   & 0.03$^a$               &                  &$>4.9$& $<$525      & 1988 & 43 & 43 & 7 & 44 \\
\hline
\end{tabular}
\newline
\end{center}
$^a$For GRS 1124--68 and GS 2000+25, BH spin estimates were reported with no errors quoted; here we assume the same errors as A0620--00 (which has a similarly low spin).
$^b$The distance and BH mass of some BHXBs are uncertain \citep[e.g.][]{hyneet04,hiemet11}; here we assume 8 kpc (since these sources are towards the galactic centre) and 10 M$_{\odot}$, respectively.
The peak radio flux density and rise times are tabulated for each radio flare over the hard-to-soft state transition reported for each source. In the sixth column, the year is that of the radio flare.
References:
(1) = \cite{cantet10};
(2) = \cite{gouet10};
(3) = \cite{kuulet99};
(4) = NM12;
(5) = \cite*{geliet01};
(6) = \cite{remimc06};
(7) = \cite{zhanet97};
(8) = \cite{ballet95};
(9) = \cite{oroset02};
(10) = \cite{shafet06};
(11) = \cite{millet09};
(12) = \cite{parket04};
(13) = \cite{oroset11b};
(14) = \cite{steiet11};
(15) = \cite{hannet09};
(16) = \cite{hiemet11};
(17) = \cite{calvet09};
(18) = \cite{hjelru95};
(19) = \cite*{greeet01};
(20) = \cite*{rupeet05};
(21) = \cite{hyneet04};
(22) = \cite{shapti09};
(23) = \cite{koledo10};
(24) = \cite{gallet04};
(25) = \cite*{steiet12};
(26) = \cite{mcclet09};
(27) = \cite{millet12};
(28) = \cite{shapet10};
(29) = \cite{reiset11};
(30) = \cite{yanget10};
(31) = \cite{fendet99};
(32) = \cite*{greiet01};
(33) = \cite{mcclet06};
(34) = \cite{blumet09};
(35) = \cite{rodret95};
(36) = \cite{rushet10};
(37) = \cite{trusni10};
(38) = \cite{reidet11};
(39) = \cite{oroset11a};
(40) = \cite{gouet11};
(41) = \cite{fabiet12};
(42) = \cite{fendet06};
(43) = \cite*{barret96};
(44) = \cite{hjelet88}.
\end{table*}

\section{The sample}

In Table 1 we list all BHXBs that we are aware of with both BH spin (measured either by the continuum fitting method or the reflection fitting method) and observed transient jet flares. BH spin and radio data are taken from table 1 of FGR10, table 1 of \cite{fendet04} and recent estimates noted in NM12, plus some new BH spin estimates published since FGR10 but not used in NM12 and radio flare data from the same BHXBs (see Table 1 for all original references for the BH spin and radio data, as well as distances and BH mass estimates). Where more than one spin estimate exists using the same method for the same BH, we use the most recent spin measurement, inferred using data from modern X-ray telescopes, since these are likely to invoke the most sophisticated fitting techniques and use the highest quality data.

\subsection{BH spin parameters}

Throughout this paper we consider both leading methods of measuring BH spin, and do not favour one method over the other. It is not in the scope of this paper to argue that one method provides more accurate spin estimates than the other; a scientifically fair study should consider all possible published measures. In the long term, the hope is that the two methods will converge and give consistent results. This appears to be currently the case for three BHXBs \citep[GRS 1915+105, Cyg X--1 and XTE J1550--564;][]{mcclet06,blumet09,steiet11,gouet11,fabiet12}.

For some BHXBs, multiple spin measurements using the same method have been broadly consistent with each other, whereas for other BHXBs they have not. Two examples are the various spin estimates based on the disc continuum fitting method for GRO J1655--40; $a_* \sim 0.93$ \citep{zhanet97}, $a_* = 0.78 \pm 0.10$ \citep*{gieret01}, $a_* < 0.7$ \citep{davihu06}, $a_* = 0.7\pm 0.1$ \citep{shafet06}, and the reflection fitting method for Cyg X--1; $a_* = 0.05 \pm 0.01$ \citep{millet09}, $a_* = 0.88^{+0.07}_{-0.11}$ \citep{duroet11}, $a_* = 0.97^{+0.014}_{-0.02}$ \citep{fabiet12}. In almost all cases the most recent measurement is likely to be the most accurate one, but we caution that this may not always be the case, and it depends on the many subtleties of the techniques and assumptions adopted in each individual case, which we cannot go into here. Generally speaking, for most BHXBs the error bars of most, if not all measurements overlap when one spin measurement method is considered, but in a few cases there are large differences (like that of Cyg X--1 mentioned above; see also the discussion in FGR10).

\subsection{Jet power from radio flares}

Since the peak flux of the bright radio flare of a BHXB can differ dramatically for different outbursts of the same source, we tabulate the brightest radio flare observed during state transition for each outburst of each source. In addition to the sources considered here, there are several other notable cases of XBs with a large range in peak radio flare fluxes \citep*[e.g. Cyg X--3, Cir X--1, SS 433;][]{preset83,spenet86,bonset86,miodet01,tudoet08,koljet10,willet11,blunet11,corbet12}.

For one source, 4U 1543--47, a radio flare seen during the state transition during its 2002 outburst was one of the few in the literature where the rise of the flare was reported and well sampled in time \citep{parket04}, rising at 0.8--1.0 GHz from $6 \pm 2$ mJy to $16 \pm 2$ mJy in 6.0 h, then peaking at $21.9 \pm 0.6$ mJy just 2.0 h later. Less than a day (19.0 h) after this brightest detection, the source had faded to $< 2.7$ mJy at 0.8 GHz. The light curve morphology is typical of a flare event often seen over the state transition \citep[see e.g.][]{gallet04}. Radio flares typically rise faster than they fade, so it is unlikely that the peak of this radio flare was much brighter than the observed peak of 22 mJy (NM12 take this flux density as a lower limit). In addition, the rise times of all radio flares are consistent with being around 5 -- 12 h \citep[Table 1; see also table 1 of][]{fendet04}. As an extremely conservative estimate, the peak could have been twice as bright as that observed, so we estimate a peak radio flux density of 22--44 mJy at 0.8--1.0 GHz. We use the method of NM12 of assuming $\alpha = -0.4$ for the spectral index of the optically thin radio flare in order to estimate the 5 GHz flux, which was 11.6--23.4 mJy.

GRO J1655--40 had a very bright radio flare of 2 Jy in 1994 which was associated with relativistic ejecta \citep{hjelru95,tinget95}, whereas in 2005, an optically thin radio flare was detected at $6.6 \pm 0.4$ mJy, $3.6 \pm 0.1$ mJy and $3.8 \pm 0.3$ mJy at 1.43 GHz, 4.86 GHz and 8.46 GHz, respectively, which then decayed over the next few days \citep{rupeet05}. The radio light curve is well sampled, with daily coverage \citep*[see also][and references therein]{shapet07,fendet09} and the rise and decay of the flare were observed. This was the brightest flare seen from GRO J1655--40 during the hard to soft state transition of this outburst, and occurred within days of the source entering the soft state \citep{rupeet05,shapet07}, which is fairly typical \citep{fendet09}. From these two outbursts with well sampled radio monitoring, it is clear that the peak luminosity of the flare associated with the state transition can vary by three orders of magnitude. The fainter flare in 2005 had a longer rise time, and the resulting total energy required to produce the 1994 and 2005 radio flares may only differ by only a factor $\sim 4$ (see Section 3).

\begin{figure*}
\centering
\includegraphics[width=8.0cm,angle=270]{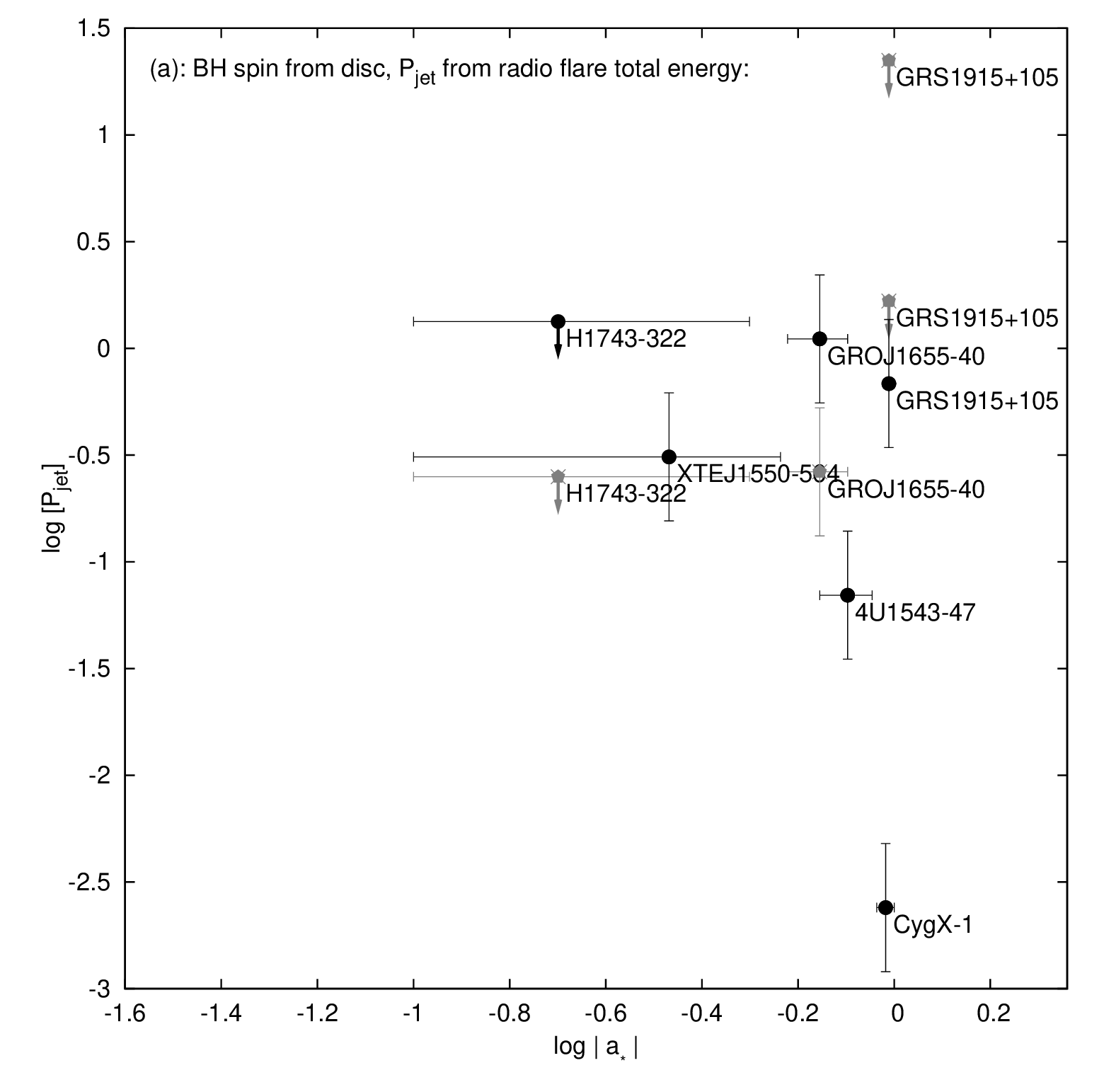}
\includegraphics[width=8.0cm,angle=270]{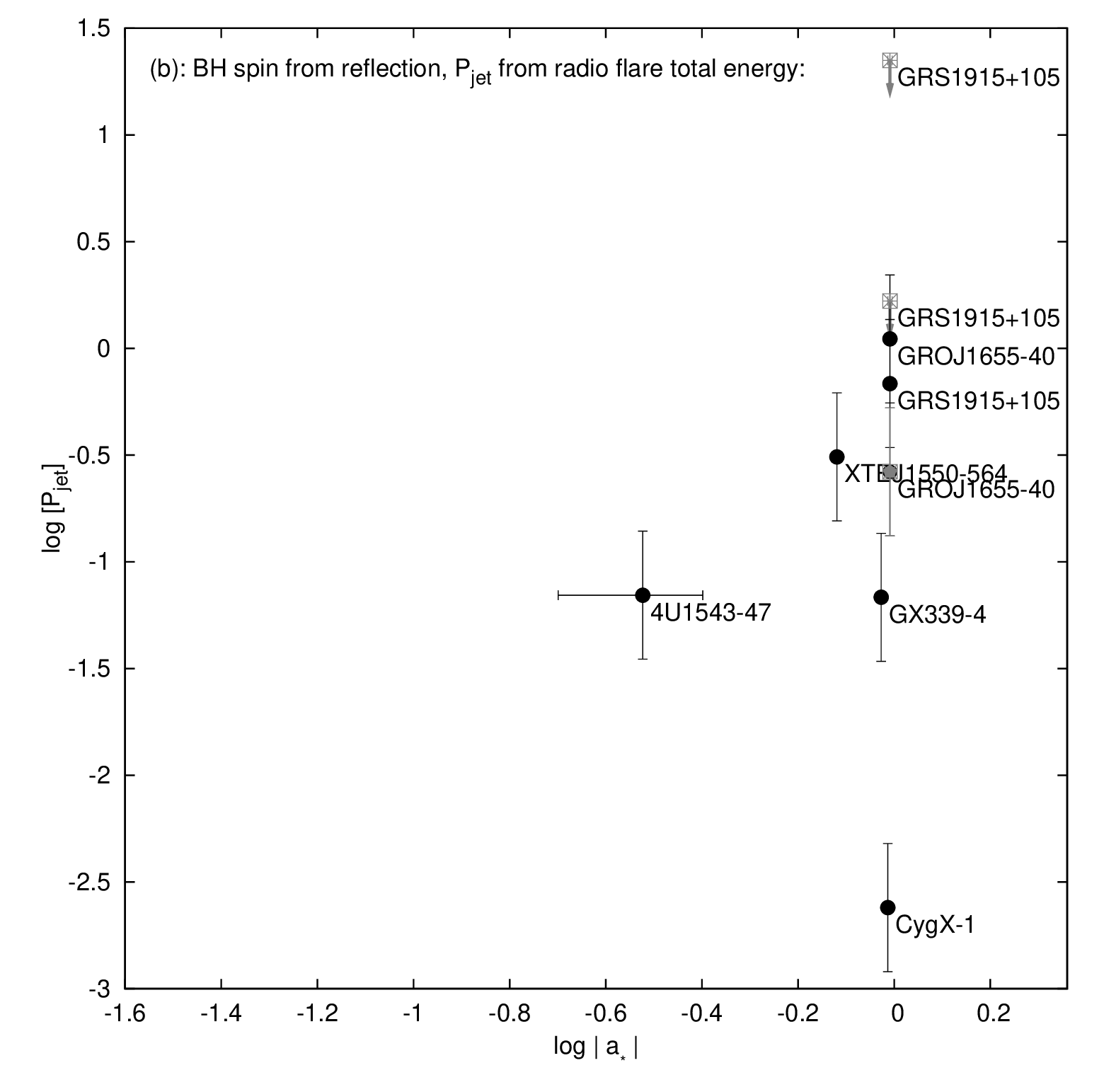}\\
\includegraphics[width=8.0cm,angle=270]{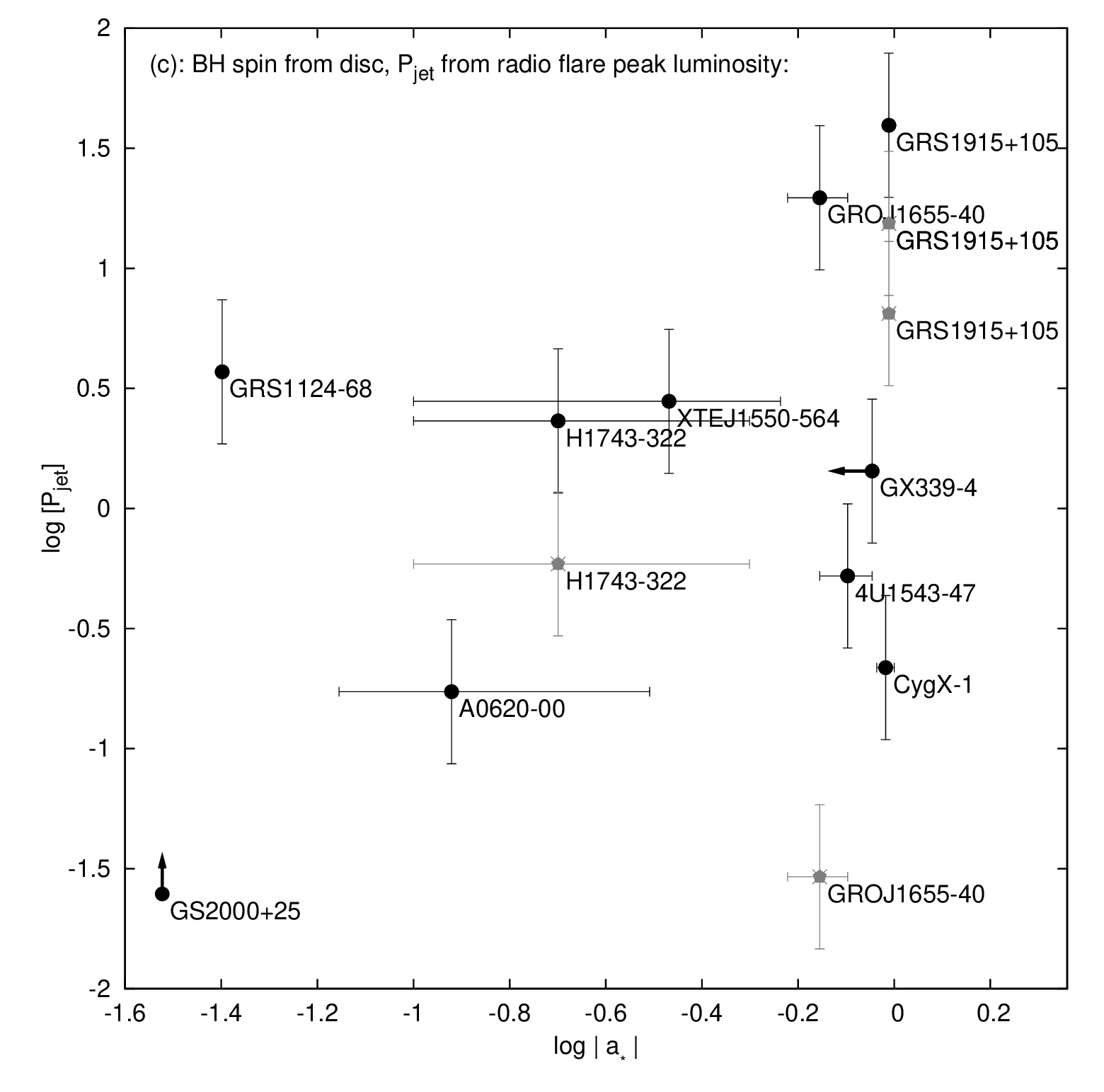}
\includegraphics[width=8.0cm,angle=270]{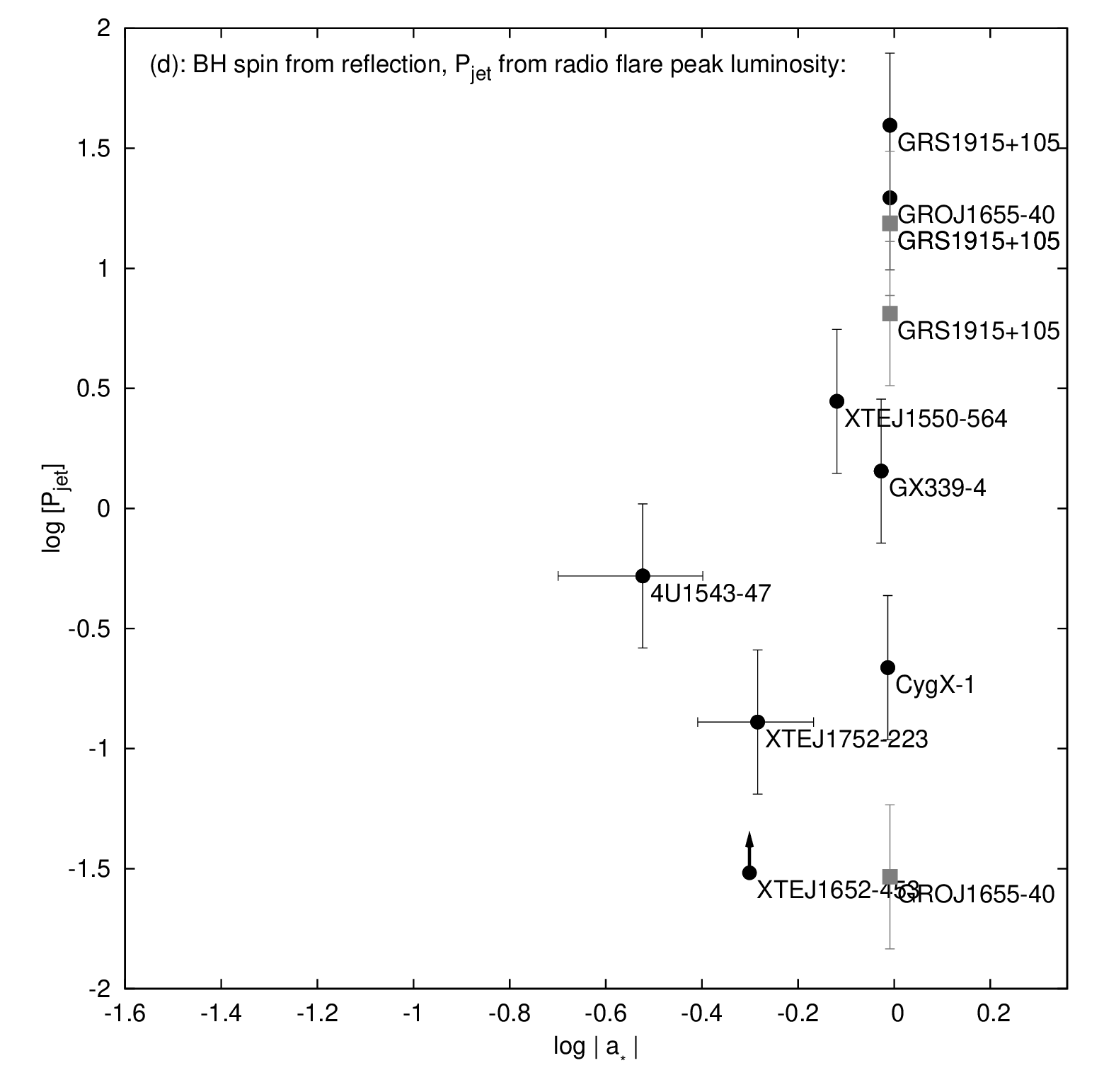}
\caption{BH spin versus relative jet power. In the left panels BH spin measured via the disc continuum method are plotted; in the right panel BH spin measured via the reflection method are plotted. The jet powers estimated from total energy (FGR10) and peak radio luminosity (NM12) and shown in the upper and lower panels, respectively. NM12 and \citeauthor{steiet13} (\citeyear{steiet13}) used a subsample of the data shown in panel (c).}
\end{figure*}

Four bright 5 GHz radio flares of GRS 1915+105 are known to have peaked at flux densities between 150 and 912 mJy (Table 1). NM12 already noted that the bright radio flare of GRS 1915+105 seen in 1997 peaked at a luminosity $\sim 3$ times fainter than the bright radio flare in 1994, which invalidates a single relation between BH spin and peak radio flare luminosity because the BH spin cannot change in three years (NM12 only included the brightest of these two flares of GRS 1915+105 in their following analysis of the BH spin--jet power relation). \cite{rushet10} found from high resolution radio imaging of GRS 1915+105 that even fainter radio flares are resolved into relativistic discrete ejecta. We therefore include all four bright radio flares in Table 1 and the following analysis. These bright radio flares are not to be confused with the fainter, quasi-regular `oscillation' flares which peak at $\sim 50$ mJy \citep[e.g.][]{fendet04}, and which are not included here (but which could be due to the same physical process). H1743--322 also had peak radio flare flux densities in 2003 and 2009 that differed by a factor of $\sim 4$ \citep{mcclet09,millet12}.

We also include Cygnus X--1 in our analysis. NM12 discard this source because, as they argue, the mass transfer is via a stellar wind from its high-mass companion star. Although Cyg X--1 is a wind-fed system, an accretion disc has been detected around the BH, its size has been estimated \citep*{coriet12}, and the disc has even been used to infer the BH spin \citep[e.g.][]{millet09,gouet11}. On large scales, the mode of accretion for wind-fed systems differs to that of Roche lobe overflow systems -- the stellar wind ensures a relatively steady, high mass accretion rate onto the outer accretion disc. The influence of BH spin on jet production is only important on distances of $\ll 100 R_{\rm g}$ from the BH in which general relativistic frame dragging becomes important \citep[e.g.][]{meieet01,mckiet12}. Obviously the mode of accretion towards the BH on these much smaller size scales is the same for wind-fed systems as (transient) Roche lobe accreting systems (i.e., matter is accreted from the inner disc into the inner regions that are affected by frame dragging), and is not affected by the mode of accretion on large scales (accretion disc or wind-fed accretion).

In addition, it is well known that Cyg X--1 performs state transitions that are associated with radio flares \citep[e.g.][]{fendet06,wilmet07,rushet12} like other BHXBs. The brightest radio detection of Cyg X--1 ever seen over years of detailed, long-term monitoring with the Ryle and Green Bank telescopes \citep[e.g.][]{brocet99,lachet06} was during the hard to soft state transition in 2004, where the radio luminosity peaked at 120 mJy \citep[][in addition, a likely extended, discrete ejection has been seen]{fendet06}. In fact Cyg X--1 spends 10 per cent of its time in the soft state \citep[e.g.][]{gallet05} in which the radio emission is suppressed \citep[e.g.][]{tigeet04,rushet12}, as expected after crossing the `jet line' when the ballistic ejection is usually launched \citep{fendet04}. Cyg X--1 should therefore certainly be included; here we use the aforementioned brightest radio flare of 120 mJy at 15 GHz. Again adopting $\alpha = -0.4$ (NM12) we estimate a 5 GHz flux density of 186 mJy.

\section{Analysis and Results}

The two methods we adopt to estimate the relative jet power of the discrete ejections are explained in FGR10 and NM12, respectively. In FGR10 we adopted the classical calculation of the minimum total energy required to energise a synchrotron flare, $P \propto (c \Delta t)^{9/7} L^{4/7}$ where $L$ is the peak radio luminosity \citep{burb59,long94}. This calculation is frequently used in the study of synchrotron radiation from numerous sources at wavelengths from radio to X-ray, and is a reliable measure the minimum total luminosity of an individual flare. We calculate the jet power of each radio flare listed in Table 1 as a fraction of the Eddington luminosity (using the known values of distance and BH mass). The second method of measuring jet power is simply using the luminosity at the peak of each radio flare (NM12), which assumes the peak luminosity is a linear indicator of total jet kinetic power \citep[contrary to the generally accepted theoretical works; e.g.][]{blanko79,falcbi95,market03,heinsu03}. This method contains fewer uncertainties, but does not take into account the different temporal profiles and hence different total energies of each radio flare. Note that for both methods, the jet power as a fraction of the X-ray luminosity is not considered, because quasi-simultaneous X-ray luminosities were not available for all radio jet flares.

In Fig. 1 we plot BH spin versus relative jet power for all data listed in Table 1. We assume an error of a factor of two for the jet power estimates. For those sources for which different measurements/limits of the jet power have been reported over different epochs, the highest values correspond to filled black markers, while grey markers correspond to lower values/limits. It is clear that there is a large scatter in the values of jet power for a single source, because transient jets had different peak fluxes and total energies during different outbursts.

BH spins of $a_{\ast} \geq 0.7$ measured from disc continuum fitting have a range of jet flare powers spanning $\geq 2.5$ orders of magnitude (left panels of Fig. 1). With the inclusion of all the available data, the claimed $P_{\rm jet} \propto a_{\ast}^2$ relation of NM12 is no longer as obvious (Fig. 1c). Particular sources to note are 4U 1543--47, Cyg X--1 and GRO J1655--40 (2005 outburst), which all had radio flares orders of magnitude fainter than expected from the relation, whereas the radio flare of GRS 1124--68 was two orders of magnitude too bright for its low BH spin (although we caution the reader that this is a BH spin measurement inferred from old data, which as yet has not been superseded). 

Spins of $a_{\ast} \geq 0.9$ measured via reflection have jet powers ranging $\geq 1.5$ and $\sim 3$ orders of magnitude for jet powers estimated from their total energies and peak luminosities, respectively. There is no clear single relation between jet power and BH spin in any of the four panels in Fig. 1.

\begin{figure}
\includegraphics[width=8.6cm,angle=0]{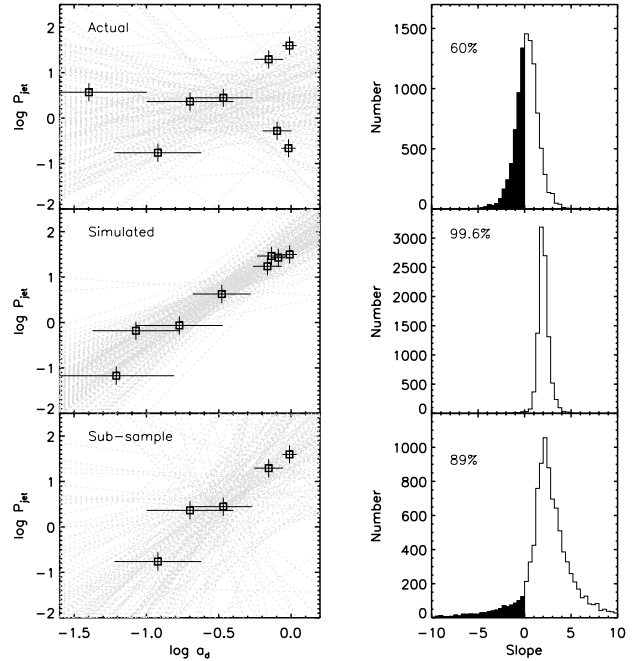}
\caption{Linear regression analysis results for spin parameter as estimated from thermal continuum vs. jet power from the peak luminosity of the radio flare (after NM12). The right column shows the posterior distribution of the fitted slope, with negative values highlighted in black. The left column shows a random selection of 150 out of the 10,000 fitting iterations that are represented to the right, over-plotted to the data with grey dotted lines. The top panels include the full data set discussed in this paper; the middle panels treat a simulated distribution of data points with the same values for the spin parameter as in the top panels, but jet powers randomly drawn from a positive correlation of the form $P_{\rm jet}\propto a_{\ast}^2$. The bottom panels repeat the same exercise as in the top panels, but for an arbitrary sub-sample of data points. The labels in the right column denote the inferred range of values for the slope of the correlation, while the percentages refer to positive slopes. Percentages lower than $\sim 95$ correspond to no statistically significant positive correlation (since more than $\sim 5$ per cent of the draws result in a \emph{negative} slope). \label{fig:bayes1}}.
\end{figure}

\subsection{Statistical analysis}

In order to quantify our visual assessment, we investigate the presence of a positive correlation between jet power and BH spin using a Bayesian code \citep{kell07}. 
We carry out a linear regression analysis on a relation of the form ${\rm log}(P_{\rm jet})=\alpha+\beta {\rm log}(a_{\ast})$, with intrinsic random
scatter included, and for the four possible permutations between the two different methods for estimating the transient jet power and and spin parameter, respectively. Three Gaussians (for the slope, intercept and intrinsic scatter) are used in the independent variable mixture modelling, and a minimum of 10,000 iterations are performed. The most likely parameter values are estimated as the median of 10,000 draws from the posterior distribution. 
For those sources for which different values of the radio flare luminosity have been reported over different epochs, we adopt {\it the highest} value as the tracer for jet power (corresponding to the filled black marks in Fig. 1).

\begin{figure*}
\centering
\includegraphics[width=16cm,angle=0]{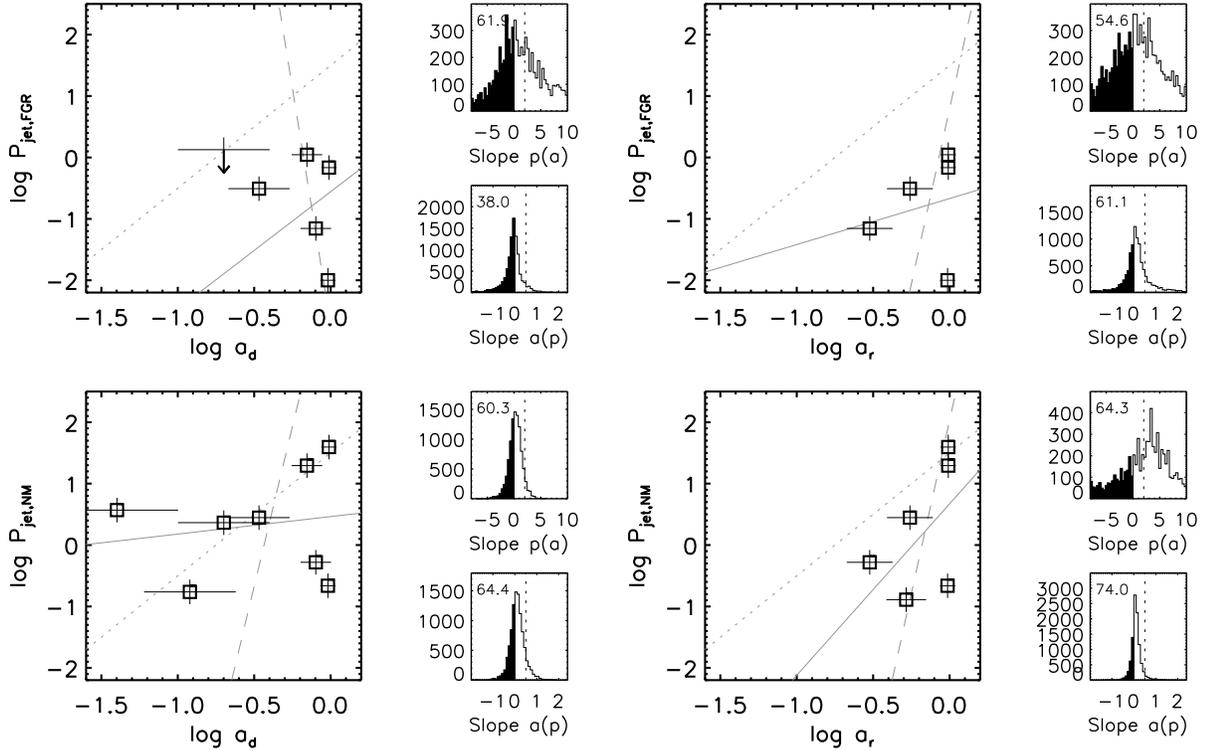}
\caption{Same as Fig. \ref{fig:bayes1}, but for different combinations of methods for estimating the spin parameter $a$ and jet power $p$. $a_{\rm r}$ and $a_{\rm d}$ denote spin parameters from reflection and disc fitting, respectively. $P_{\rm FGR}$ and $P_{\rm NM}$ refer to jet power from total energy and peak flare luminosity, respectively. The solid/dashed lines are the fitted $p(a)$, $a(p)$ slopes, compared to a $p\propto a^2$ slope (dotted line). Dotted lines in the histograms are for slopes of 2 ($p$ vs. $a$ fit) and 0.5 ($a$ vs. $p$). The numbers in the two histograms next to each fit are the percentage of draws that have slopes consistent with some positive correlation between  $a$ and $p$ being present (i.e. the unfilled fraction).  
The data points do not show evidence for a significant correlation, for any of the four $a$ versus $p$ combinations. 
\label{fig:bayes2}
}
\end{figure*}

The top panels of Fig. \ref{fig:bayes1} illustrate the results of the linear regression analysis for the case where a correlation has been claimed, i.e. for the jet power as inferred from the peak luminosity of the radio flare versus the spin parameter as estimated from the thermal continuum fitting method (NM12), here denoted as $a_{\rm d}$. The histogram on the top right shows the distribution of the fitted slopes, with the draws returning a negative slope highlighted in black. For this specific case, only 60 per cent of the draws have slopes consistent with a positive correlation between the spin parameter and and jet power being present, {\it ruling out a positive correlation with high confidence} ($\simgt 5~\sigma$). This is perhaps most apparent in the top left panel, where a random sub-sample of 150 fits (out of 10,000) is over-plotted to the data. 

For comparison, the middle panels correspond to a similar analysis as the one discussed above, only for a {\it simulated} data set where the spin parameter values (along with errors) are identical to those in the top panels, while the jet power values are randomly drawn from a $P_{\rm jet} \propto a_{\ast}^2$ distribution with an intrinsic scatter of 0.5 dex (additionally, each jet power value is assigned an error of 0.3 dex).  In this case, only 0.4 per cent of the draws return negative slopes, which indicates a statistically significant ($\simgt 3~\sigma$) positive correlation. By comparison with the top panels, this exercise serves to illustrate that the lack of a statistically significant correlation in the full data set is not merely a result of the small number statistics, and the code could indeed recover a statistically significant correlation with eight data points only. 

Lastly, the bottom panels illustrate how the results of the regression analysis change for five out of the eight data points considered in the top panels (specifically with GRS1124$-$68, Cyg X$-$1 and 4U~1543$-$47 excluded from the analysis; as in NM12). The sample is actually the same as that in \cite{steiet13}; this is discussed further in Section 4. While the slope distribution is formally consistent with a value of two, the significant tail of negative draws (11 per cent) corresponds to a marginally significant correlation only ($\simlt 2~\sigma$), although it is worth stressing that the statistical significance of the correlation for this data set is very sensitive to the error bars (the Bayesian code we are adopting can only treat with symmetric error bars. 

The same analysis as described above was carried out to assess the presence of a relation between the spin parameter $a$ and the jet power $p$ as measured from the different methods, namely spin from reflection ($a_{\rm r}$) and jet power from total energy (denoted as $P_{\rm FGR}$, vs. $P_{\rm NM}$). Plots analogous to those discussed in Fig. \ref{fig:bayes1} are shown in Fig. \ref{fig:bayes2} for the following combinations of parameters: jet power from peak radio flare luminosity vs. spin parameter from thermal continuum (top left); jet power from peak radio flare luminosity vs. spin parameter from reflection (bottom right); jet power from flare total energy vs. spin parameter from thermal continuum (top left); jet power from flare total energy vs. spin parameter from reflection (top right). As apparent from the prominent negative tails of the fitted slope distributions, no positive, statistically significant correlation is found in all cases. A similar conclusion is reached by fitting either $p$ as a function of $a$ (solid lines), or $a$ as a function of $p$ (dashed lines).

\section{Discussion and summary}

We have shown that no single statistically significant relation exists between inferred transient jet power and BH spin parameter, with the current estimates of these parameters. Furthermore, the very fact that one BHXB can produce radio ejections with very different total energies and peak luminosities invalidates a single relation between transient jet radiative energy and BH spin. This latter discovery implies that many jet flares may be less luminous than the maximum value that would be produced by a purely spin-powered jet. For a typical source, there have been a limited number of outbursts observed to date, so it could be that truly spin-powered jets do reach a constant, maximum jet power, but few flares reach this critical maximum. This could lead to underestimated jet luminosity estimates. We note that one source, GRS 1124--68 appears to have a high jet power but a low BH spin, which is not consistent with the above picture. Regular radio monitoring of populations of BHXBs over state transitions is required to measure the distribution of jet powers of discrete ejections.

It is worth noting that in our study we do not include the effects of relativistic beaming of the jet emission. This is because both the orbital inclination angles (which are usually assumed to be perpendicular to the jet axis) and the jet Lorentz factors are uncertain for many sources (NM12 assumed a Lorentz factor of $\gamma = 2$ for all sources).

Very recently, a new paper has appeared from the same group as NM12, which again claims a correlation between BH spin (measured from continuum fitting) and jet power (measured from peak luminosities) in transient ejections \citep{steiet13}. In this work, the same subsample of BHXB spin measurements that appeared in NM12 was used, with an addition of one more spin measurement, that of H1743--322 from \cite{steiet12}. This spin measurement is also included in our analysis. However, it is worth noting a few discrepancies between our analysis and theirs.

For this source we took jet flares from two outbursts, 2003 and 2009, whereas in \cite{steiet13} just the former jet flare is used. Furthermore, we adopted the radio flare peak luminosity associated with the hard to soft transition in Fig. 1c (the same method as was adopted in NM12) but \cite{steiet13} use a fainter flare, which peaked at 35 mJy at 5 GHz instead of the brighter, 96 mJy flare we used. Their arguement is that this fainter flare must have been the one associated with the launch of the jet ejection because the date on which this ejection occurred is known from resolved radio observations. However, if the BH spin powers the 35 mJy flare but not the 96 mJy flare, then what powers the brighter flare? The 96 mJy flare had an optically thin spectrum and did not occur during the hard state \citep{mcclet09}, so we can rule out a compact jet origin to this brighter flare. If one of those two flares was associated with an energetic boost from the spin of the BH, one would expect it to be the brighter one.

Furthermore, since we showed that radio ejections observationally possess very different peak luminosities, the peak flux cannot be used as a proxy for the total power of a discrete ejection. \cite{steiet13} do not take into account the time over which a flare is radiated (which empirically can vary by more than one order of magnitude; see Table 1), but instead adopt $(\nu L_{\nu})_{\rm max} \propto P_{\rm jet}$ \citep[see equation B10 in the Appendix of][]{steiet13}, contrary to standard classical synchrotron equation, $P_{\rm jet,min} \propto (c \Delta t)^{9/7} L^{4/7}$.

Finally, a further discrepancy lies in the assumed relation between BH spin and jet power. In fitting the correlation, \cite{steiet13} assume $P_{\rm jet} \propto (a_{\ast} / (2(1+(1-a_{\ast}^2)^{1/2})))^2$, as opposed to that adopted in NM12, $P_{\rm jet} \propto a_{\ast}^2$, assume the same jet Lorentz factor for all sources, and fix the power law index to 2.0 (as indicated in footnote 6 in their paper). For comparison, the Bayesian, linear regression analysis we present here leaves the slope as a free parameter, testing the probability that it is positive at all. As we discussed in detail in Section 3, a full Bayesian treatment, with error bars on both jet power and spin parameters included, shows that a positive correlation of the form $P_{\rm jet} \propto a_{\ast}^2$ is at best marginally significant ($\simlt 2~\sigma$), even when the new H1743--322 data point is included.

If any of the above methods of measuring BH spin or jet power are correct, then we can conclude that there is no single relation between BH spin and transient jet power. 
Are there any reasonable explanations that could explain the lack of a correlation? Many works predict (analytically or via simulations) a relation between the spin of a black hole and the power of the resulting accretion-fed jet \citep*[e.g.][]{macdth82,liviet99,mcki05,deviet05,hawlkr06,tcheet10,mckiet12}. Positive relations are generally predicted but not necessarily as simple as the $P_{\rm jet} \propto a_{\ast}^2$ relation originally derived by \cite{blanzn77}. It has recently been shown that rapidly \emph{retrograde} spinning BHs (where the BH spin is in the opposite direction to the accreting matter spiralling inwards) can produce jets just as powerful as rapidly prograde spinning BHs \citep[e.g.][]{garo09,meie11,tchemc12}. Interestingly, the only source with a (slightly) negative spin in Table 1, GRS 1124--68, had a relatively powerful jet (Fig. 1; lower left panel). However, since several BHs with very similar spin estimates have produced flares of energies and peak luminosities spanning several orders of magnitude (including different jet powers for different flares from the same source) we can rule out a single relation between jet power of discrete ejections and current estimates of BH spin. This is quantified by our Bayesian analysis (Figs. 2 and 3). 

The spin parameters as estimated from the two methods do not agree with each other for a number of sources, and neither is to be preferred. In addition, for compact, steady jets that exist in the hard state, radio or infrared luminosity normalization may turn out to be an unreliable tracer for jet power. Dual tracks exist with sources jumping between them and having different correlation slopes \citep{russet07b,coriet11,gallet12}, which implies that the jet luminosity is affected by additional, as yet undefined parameters as well as the mass accretion rate and (possibly) the BH spin. The radio/X-ray (and IR/X-ray) correlation normalization therefore may not be an accurate method of estimating the relative ranking of compact jet power, however the range in jet luminosities in the hard state (for a given X-ray luminosity) could still be improved. Nevertheless, the results of FGR10 suggest there may not be a strong dependence of relative compact jet power on BH spin, even given the above caveats.

It was also predicted by \cite{blanzn77} that jet power also scales with the square of the magnetic field strength; $P_{\rm jet} \propto a_{\ast}^2 B^2$, because the mechanism relies on differential rotation between a spinning BH and its magnetosphere. One would therefore not expect a $P_{\rm jet} \propto a_{\ast}^2$ correlation to exist unless the magnetic field is nearly constant across all sources \citep[this caveat is also mentioned, but then later ignored, in][]{steiet13}. In the hard state, it may be that $B^2 \propto \dot{m}$, the mass accretion rate \citep[e.g.][]{heinsu03}, and in \cite{sikobe13} it is argued that jet power is mostly driven by the magnetic flux, not the spin or accretion rate. Jet luminosity (measured in radio or IR) does scale positively with X-ray luminosity and hence with mass accretion rate, in the hard state. However this is no indication of the Blandford \& Znajek mechanism, as a positive relation between X-ray luminosity and jet power is of course also expected for the Blandford--Payne mechanism (in which accretion powers the jet). In addition, \cite{nixoki13} have recently argued that jets that precess on short timescales cannot be spin powered.

Even if BHXB jets were powered by BH spin, the combination of several effects -- including but not limited to uncertainties in measuring the relevant parameters -- could still prevent us from finding a positive correlation at present.

\section*{Acknowledgements}

DMR would like to thank Jon Miller and Sera Markoff for insightful discussions. EG is grateful to Brendan P. Miller for his assistance with the Bayesian analysis, and to Monica Colpi and Laura Maraschi for their insights. DMR acknowledges support from a Marie Curie Intra European Fellowship within the 7th European Community Framework Programme under contract no. IEF 274805.

\label{lastpage}

\end{document}